\documentclass[prb,showpacs]{revtex4}
\usepackage{graphicx,psfrag,amssymb,amsmath,color}
\begin{document}
\title{The pseudogap phase in (TaSe$_4$)$_2$I}
\author{Bal\'azs D\'ora}
\email{dora@kapica.phy.bme.hu}
\author{Andr\'as V\'anyolos}
\affiliation{Department of Physics, Budapest University of Technology and 
Economics, H-1521 Budapest, Hungary}
\author{Attila Virosztek}
\affiliation{Department of Physics, Budapest University of Technology and 
Economics, H-1521 Budapest, Hungary}
\affiliation{Research Institute for Solid State Physics and Optics, P.O.Box
49, H-1525 Budapest, Hungary}

\date{\today}

\begin{abstract}
We have developed the mean-field theory of coexisting charge-density waves
(CDW) and unconventional charge-density waves (UCDW). The double phase transition manifests itself in the thermodynamic 
quantities and in the magnetic response, such as spin susceptibility and spin-lattice relaxation rate.
Our theory applies to quasi-one dimensional (TaSe$_4$)$_2$I, where above the CDW transition, thermal fluctuations die 
out rapidly, but  robust pseudogap behaviour is still detected. We argue, that the fluctuations are suppressed 
due to UCDW, which 
partially gaps the Fermi surface, and causes non-Fermi-liquid (pseudogap) behaviour.

\end{abstract}


\maketitle

\section{Introduction}

One dimensional interacting metals are unstable with respect to charge-density wave (CDW) formation. A 
strictly one-dimensional system cannot exhibit long-range CDW ordering above $T=0$, although strong CDW fluctuations 
occur. Mean-field theories provide us with a finite $T_c$ by neglecting the fluctuations. However, increasing the 
dimensionality of the system by weak interchain coupling can lead to long-range ordered 
CDW at transition temperature $T_c$ well below the mean-field transition temperature. Consequently, quasi-one 
dimensional systems were expected to possess CDW ground state, which was confirmed by experiments\cite{gruner}.

Among the possible candidates, inorganic conductors as NbSe$_3$, TaSe$_3$, K$_{0.3}$MoO$_3$ and (TaSe$_4$)$_2$I have 
attracted much attention. Due to their quasi-one dimensional nature, CDW ordering and fluctuation should be considered 
on 
equal footings, as was done in the pioneering work of Lee, Rice and Anderson\cite{lra}. Their results fitted a variety of 
results on the thermodynamic and magnetic properties of inorganic conductors in the fluctuation regime slightly below 
and 
above $T_c$\cite{johnston1,johnston2}. However, the last member of the materials cited above, namely (TaSe$_4$)$_2$I, 
exhibits peculiar 
properties:

1. Sharp increase of the static spin susceptibility above $T_c$ in a wide temperature range\cite{johnston2,Kriza}.

2. Smooth increase of the spin-lattice relaxation rate above $T_c$\cite{Kriza}.

3. As opposed to conventional CDW, no sharp features  in the spin lattice relaxation rate below $T_c$\cite{Kriza}.

4. Thermal fluctuations die out rapidly with increasing temperature above $T_c$\cite{requardt}.\\
The first has been interpreted within the fluctuating gap model\cite{johnston2}, whose application is questionable in 
light of the forth. The other two has not been addressed theoretically so far.

Instead of using the fluctuating gap model, we propose an alternative model: the low temperature phase of 
(TaSe$_4$)$_2$I is dominated by CDW, but the response of the high temperature or pseudogap phase is attributed to 
unconventional 
charge-density waves\cite{mplb} (UCDW). Unconventional charge density waves are density waves with a wavevector 
dependent gap, whose average vanishes on the Fermi surface ($\langle\Delta({\bf k})\rangle=0$). As a result, no 
modulation of the charge or spin density is expected in these systems, and the notion "hidden-order" applies naturally. 
The gap function has zeros at 
the Fermi surface, where low energy excitations, characterized by Dirac fermions, occur. The thermodynamic and 
transport properties of unconventional density waves have been worked out, for a review see Ref [\onlinecite{mplb}].
Such phases were proposed for various materials, e.g. for the pseudogap phase in high $T_c$ 
superconductors\cite{nayak,carbotte}.

In the followings, we are going to discuss the coexistence of CDW and UCDW. Within the mean-field approximation, we 
determine  the phase diagram, and find the region of coexistence. Interestingly, a CDW gap can open in an existing UCDW 
phase, but the opposite cannot occur. The spin susceptibility and spin lattice relaxation rate keep on increasing when 
raising the temperature above the CDW $T_c$, in accordance with experimental data\cite{Kriza}. In addition, no critical 
divergence of the relaxation rate below 
$T_c$ is found, as opposed to the usual CDW response\cite{maniv}. Whence, the four basic properties of the 
pseudogap phase are satisfied by our CDW+UCDW model. For the first three, direct calculations are shown to evidence the 
agreement, while the last is argued to be obeyed since the number of normal particles is heavily suppressed above CDW 
due to the presence of UCDW condensate. 
To explore the features of our model in full glory, further studies are needed.

\section{Thermodynamics, phase diagram}

\begin{figure}[h!]
\psfrag{x}[t][b][1.2][0]{$P_0/P_1$}
\psfrag{y}[b][t][1.2][0]{$T/T_{c1}$}
\psfrag{n}[t][b][1.2][0]{normal state}
\psfrag{u}[t][b][1.2][0]{UCDW}
\psfrag{c}[t][b][1.2][0]{CDW}
\psfrag{c+u}[][][1.2][0]{{\color{red}CDW+UCDW}}
{\includegraphics[width=7cm,height=7cm]{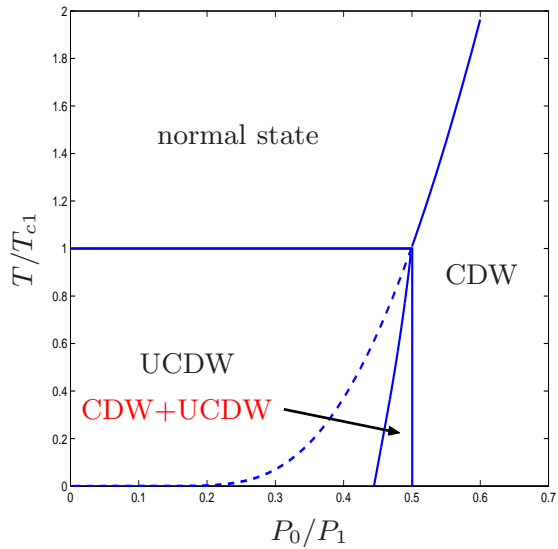}}
\caption{(Color online) The mean-field phase diagram is drawn for UCDW coupling
$P_1\rho_0=1$ with transition temperature $T_{c1}$, $T_{c1}/W\approx 0.0202$. The
dashed line stands for the pure CDW transition in the absence of UCDW. The transition
along solid lines is of second order. Similar phase diagram is obtained for other values of $P_1$. 
\label{phase}}
\end{figure}

\begin{figure}[h!]
\psfrag{x}[t][b][1.2][0]{$T/T_{c1}$}
\psfrag{y}[b][t][1.2][0]{$\Delta_i/\Delta_1(0)$}
{\includegraphics[width=7cm,height=7cm]{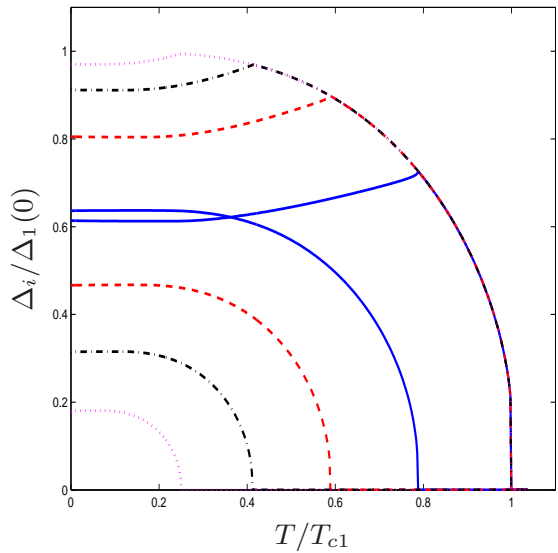}}
\caption{(Color online) The temperature dependence of the gaps are plotted for various interaction strength: $P_1\rho_0=1$ and 
$P_0\rho_0=0.49$ (blue solid line), $0.48$ (red dashed line), $0.47$ (black dashed-dotted line) and $0.46$ (magenta 
dotted line). The gaps are normalized by the gap maximum of a pure UCDW at $T=0$. The curves starting at the highest 
$T_c$ 
belong to $\Delta_1$.
\label{egy}}
\end{figure}

We consider the simple model Hamiltonian describing density waves given by\cite{gruner,nagycikk}:
\begin{equation}
 H=\sum_{\bf k,\sigma}^\prime\left[\xi({\bf k})(a_{\bf k,\sigma}^{+}a_{\bf
 k,\sigma}-a_{\bf k-Q,\sigma}^{+}a_{\bf k-Q,\sigma})+\Delta({\bf k})a_{\bf k,\sigma}^{+}a_{\bf
 k-Q,\sigma}+\overline{\Delta({\bf k})}a_{\bf k-Q,\sigma}^{+}a_{\bf k,\sigma}\right]
\label{hamilton}
\end{equation}
where $a_{\bf k,\sigma}^{+}$ and $a_{\bf k,\sigma}$ are, respectively,
the creation and annihilation operators of an electron of momentum $\bf k$ and
spin $\sigma$. In a sum with prime $k_x$ runs from $0$ to $2k_F$ ($k_F$ is the Fermi wavenumber), ${\bf 
Q}=(2k_F,\pi/b,\pi/c)$ is the best nesting vector. 
Our system is based on an orthogonal lattice, with lattice constants $a,b,c$
toward direction $x,y,z$. The system is anisotropic, the quasi-one-dimensional
direction is the $x$ axis.
The linearized kinetic-energy spectrum of the Hamiltonian is:
\begin{equation}
\xi({\bf k})=v_F(k_x-k_F)-2t_{b}\cos(k_{y}b)-2t_{c}\cos(k_{z}c)-\mu.
\end{equation}
$\Delta({\bf k})$ is the density wave order parameter, and is 
determined from the 
self-consistency condition:
\begin{equation}
\Delta({\bf l})=\sum_{\bf k,\sigma}^\prime P({\bf k,l})\langle a^+_{\bf k,\sigma}a_{\bf k-Q,\sigma}\rangle,
\end{equation}
where $P(\bf k,l)$ is a linear combination of the interaction
matrix elements, in which we considered at most nearest neighbour interactions\cite{nagycikk}.
Since we are focusing on the coexistence of CDW+UCDW, we take the kernel of the form $P({\bf 
k,l})=-P_0-P_1\sin(bk_y)\sin(bl_y)$, with both $P_0$, $P_1>0$, using the
analysis of the electron-electron interaction in quasi-one dimensional
systems\cite{nagycikk}. The sign of the matrix elements assures
charge-density wave formation. Had we chosen a gap with cosine function or with the $z$ component of the 
wavevector would not alter our results. Based on this kernel, the gap is given by
\begin{equation}
\Delta({\bf k})=\Delta_0+\Delta_1e^{i\phi}\sin(bk_y).
\end{equation}
Here we have skipped the overall phase of the condensate since it turns out to be irrelevant for most of our discussion.
The value of the relative phase of the gap coefficients, $\phi$ should be determined by minimizing the free energy of 
the system with respect to it. It turns out, that coexisting solution can occur only for $\phi=0$ 
or $\pi/2$ 
(by adding integer$\times \pi$ describes the same physics). Among these, the former solution has always higher energy than the pure 
phases, namely CDW with $\Delta_0\neq 0$ and $\Delta_1=0$ or UCDW with $\Delta_0=0$ and $\Delta_1\neq 0$. The latter 
provides us with energetically stable coexistence, and the gap now is of the form: $\Delta({\bf k})=
\Delta_0+i\Delta_1\sin(bk_y)$.
The coupled gap-equations are obtained as
\begin{gather}
\Delta_0=\rho_0P_0\Delta_0\int\limits_0^{\pi/2}\frac
{2\textmd{d}(bk_y)}{\pi}\int\limits_0^W{\textmd{d}\xi}\dfrac{\tanh\left(\beta
E/2\right)}{2E},\label{gapcdw}\\
\Delta_1=\rho_0P_1\Delta_1\int\limits_0^{\pi/2}\frac 
{2\textmd{d}(bk_y)}{\pi}\int\limits_0^W{\textmd{d}\xi}\sin^2(bk_y)
\dfrac{\tanh\left(\beta E/2\right)}{2E}\label{gapucdw},
\end{gather}
where 
\begin{equation}
E({\bf k})=\sqrt{\xi^2({\bf k})+\Delta_0^2+\Delta_1^2\sin^2(bk_y)}
\end{equation}
is the energy spectrum, $W=v_Fk_F$ is half of the bandwidth, $\rho_0$ is the
density of states in the normal state per spin at the Fermi energy. This spectrum
implies a finite energy gap as long as $\Delta_0$ remains finite. 
The phase diagram is depicted in Fig. \ref{phase}. 
Coexistence region is found only around $2P_0\lesssim P_1$, because this implies
that the transition temperature to CDW or UCDW is close to each other,
hence the two phases can compete. 
The gap-equations have been solved 
numerically and the result is shown in Fig. \ref{egy}. The system at low
temperatures can enter into a CDW or UCDW or CDW+UCDW phase. In the
following we are focusing on the latter region of
coexistence. By lowering the
temperature, first the UCDW gap opens at the transition temperature 
$T_{c1}=2\gamma W\exp(-4/P_1\rho_0)/\pi$ given by 
the weak-coupling solution\cite{nagycikk} of
Eqs. \ref{gapcdw}-\ref{gapucdw}, $\gamma=1.781$ is the Euler constant. It follows the $T$ dependence of the pure 
solution until the CDW gap opens at $T_{c0}$. Then $\Delta_1$ slightly decreases and saturates to a finite value while 
$\Delta_0$ 
grows in a mean-field manner with decreasing $T$. The opposite scenario,
the opening of UCDW gap upon an existing CDW one is excluded within the
present model, because no such solution of Eqs. \eqref{gapcdw}-\eqref{gapucdw} was found.
At $T=0$, the coexistence region is located to $4/(8+P_1\rho_0)<P_0/P_1<1/2$. 
The larger $P_1$, the wider this region is.
In the coexistence region, the amplitude of the charge-density oscillations is determined solely by $\Delta_0$, while in 
the pure UCDW phase, no modulation is found.

\begin{figure}[h!]
\psfrag{x}[t][b][1.2][0]{$T/T_{c1}$}
\psfrag{y}[b][t][1.2][0]{$C(T)/\gamma_s T$}
{\includegraphics[width=7cm,height=7cm]{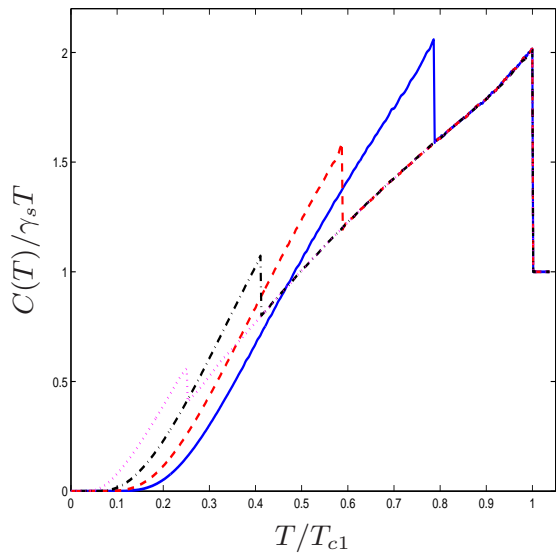}}
\caption{(Color online) The temperature dependence of the specific heat is shown for $P_1\rho_0=1$ and
$P_0\rho_0=0.49$ (blue solid line), $0.48$ (red dashed line), $0.47$ (black dashed-dotted line) and $0.46$ (magenta 
dotted line). The curves are 
normalized by the the normal state $T$ linear specific heat, $\gamma_s$ is the Sommerfeld coefficient.
\label{fajho}}
\end{figure}

Since within the mean-field theory, we are dealing with non-interacting quasiparticles, their entropy  
can be calculated from the knowledge of the energy spectrum and yields to 
\begin{equation}
S(T)=-2\sum_{\bf k,\sigma}^\prime\left(f\ln(f)+(1-f)\ln(1-f)\right),
\end{equation}
where $f=1/(\exp(\beta E({\bf k}))+1)$ is the Fermi function. By taking its derivative with respect to temperature, the specific heat can be 
evaluated.
This double phase transition is manifested in the specific heat as well, which exhibits two  peaks at the 
transitions to 
UCDW and CDW, as can be seen in Fig. \ref{fajho}

\section{Magnetic response}

\begin{figure}[h!]
\psfrag{x}[lb][rt][1.2][0]{$T/T_{c1}$}
\psfrag{y}[rb][lt][1.2][0]{$\omega/\Delta_0$}
\psfrag{z}[b][t][1.2][0]{$N(\omega)/\rho_0$}
\includegraphics[width=8cm,height=8cm]{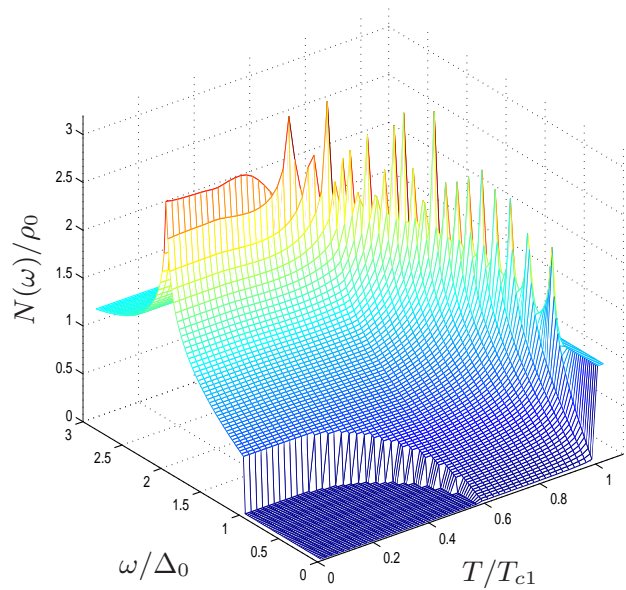}
\caption{(Color online) The temperature and frequency dependence of the quasiparticle density of states is shown for $P_1\rho_0=1$ 
and $P_0\rho_0=0.48$. The temperature dependence enters through $\Delta_i(T)$. 
\label{dos3d}}
\end{figure}

The magnetic response of the model studied in the followings includes the
long wavelength static spin susceptibility and nuclear spin-lattice
relaxation rate. These are related to the single particle density of
states, hence we begin with the exploration of its properties. We find
\begin{equation}
\frac{N(E)}{\rho_0}=2\sum^\prime_{\bf k}\delta\left(E-E({\bf
k})\right)=\frac
2\pi\int\limits_0^{\pi/2}\textmd{Re}\dfrac{|E|\textmd{d}y}{\sqrt{E^2-\Delta_0^2-\Delta_1^2\sin^2(y)}}=
N_{CDW}\left(\frac
{E}{\Delta_0}\right)N_{UCDW}\left(\frac{\sqrt{E^2-\Delta_0^2}}{\Delta_1}\right)
\end{equation}
per spin direction. Here $N_{CDW}(x)=|x|/\sqrt{x^2-1}\Theta(|x|-1)$ is the density of
states in a conventional DW,
$N_{UCDW}=\Theta(1-|x|)2xK(x)/\pi+\Theta(|x|-1)2K(1/x)/\pi$ is the density
of states in UDW. Whence, the total density of states in the coexistence
region is characterized by a
finite energy gap $\Delta_0$ stemming from the appearance of CDW, and a
logarithmic singularity at $\pm\sqrt{\Delta_0^2+\Delta_1^2}$ reminiscent to
that in UCDW. As soon as $\Delta_1$ disappears, this turns into a square-root divergence. These features can be observed 
in Fig. \ref{dos3d}.

\begin{figure}[h!]
\psfrag{x}[t][b][1.2][0]{$T/T_{c1}$}
\psfrag{y}[b][t][1.2][0]{$\chi(T)/\chi_0$}
{\includegraphics[width=7cm,height=7cm]{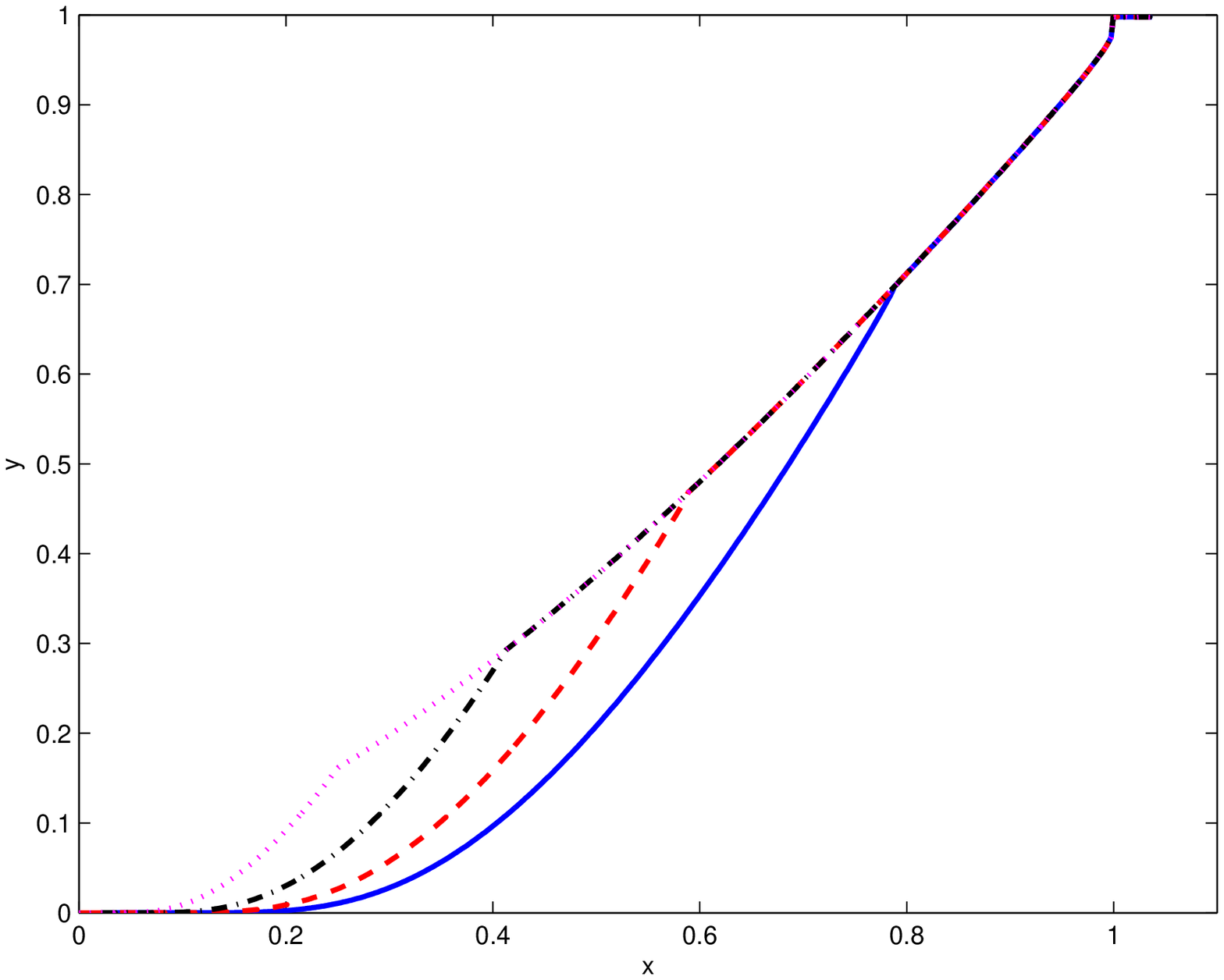}}
\hspace*{1cm}
\psfrag{x}[t][b][1.2][0]{$T/T_{c1}$}
\psfrag{y}[b][t][1.2][0]{$R(T)/R_0$}
\includegraphics[width=7cm,height=7cm]{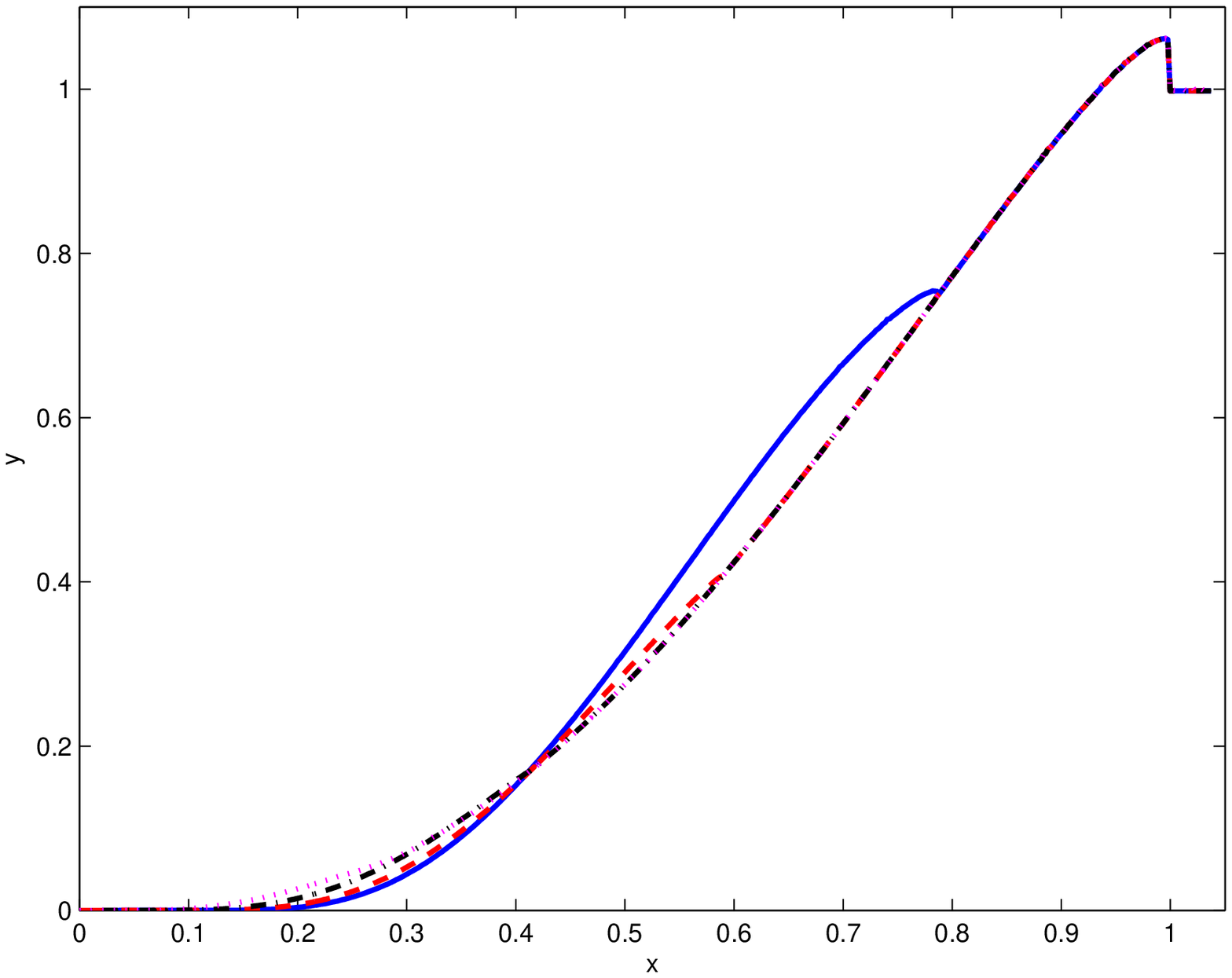}
\caption{(Color online) The temperature dependence of the spin susceptibility is shown in the left panel for 
$P_1\rho_0=1$ and
$P_0\rho_0=0.49$ (blue solid line), $0.48$ (red dashed line), $0.47$ (black dashed-dotted line) and $0.46$ (magenta
dotted line). The right panel shows the spin-lattice relaxation rate. Here the deviation for various $P_0$'s is 
smaller than 
for the spin susceptibility.
\label{spins}}
\end{figure}

From this, the spin susceptibility and the spin-lattice relaxation rate are calculated as
\begin{gather}
\dfrac{\chi(T)}{\chi_0}=-\int\limits_{-\infty}^\infty \dfrac{N(E)}{\rho_0}\dfrac{\partial f}{\partial 
E}\textmd{d}E,\label{ss}\\
\dfrac{R(T)}{R_0}=-\int\limits_{-\infty}^\infty \dfrac{N^2(E)}{\rho_0^2}\dfrac{\partial f}{\partial 
E}\textmd{d}E\label{slrr},
\end{gather}
where $\chi_0$ and $R_0$ are the respective metallic state values. The first expression applies to various kinds of 
condensates. The latter one is determined from the short wavelength spin response ($\bf q\sim Q$), because dominant 
contribution comes from wavevectors close to the ordering (nesting) vector. This explains\cite{maniv,tesla} the absence 
of coherence 
factors even in the presence of a constant gap as opposed to the type II coherence factors in 
superconductors\cite{tinkham}.
By inserting the solution of the gap equation into Eqs. \eqref{ss}-\eqref{slrr}, we obtain the temperature dependence of 
these quantities. At low temperatures, activated behaviour is seen because of the finite CDW gap. Upon increasing $T$, a 
change of slope is exhibited at $T_{c0}$, and then the curves follow that in a pure UCDW.
In the spin-lattice relaxation rate, a small peak is observable right below $T_{c1}$. In Maniv's original 
expression\cite{maniv} for $R(T)$ in a conventional DW, the magnitude of the peak was controlled by 
$\Delta\ln(8\Delta/\omega_L)$, $\omega_L$ being the Larmor frequency. In our case, the divergence in the density of 
states is 
logarithmic, hence the Larmor frequency can be taken to zero. Consequently, no sharp features are observed close to 
$T_{c0}$ as seen in Fig. \ref{spins}.
Nevertheless, a small peak remains right below 
$T_{c1}$ as 
a reminder to 
the 
density wave nature of the condensate. 

\section{Application to (T$\textbf{a}$S$\textbf{e}_4$)$_2$I}

In mean field theories, attractive interaction on lattices leads to CDW formation in low dimensional systems. In a
strictly one-dimensional chain, however, CDW fluctuations eliminate any long range order above $T=0$. In reality, finite 
interchain coupling increases dimensionality and restores long range order
at a temperature, which is an appreciable 
fraction of the mean-field value.

This is why quantum fluctuations are very important in studying low dimensional systems such as K$_{0.3}$MoO$_3$, 
TaS$_3$ 
and other CDW materials. Lee, Rice and Anderson\cite{lra} investigated 
the effects of fluctuations 
on the Peierls 
transition in one dimension by taking a functional integral over variations in the order parameter. The success 
of their 
theory became apparent when applied to the aforementioned CDW materials. They found the suppression of the mean-field 
$T_c$ by  the fluctuation. Moreover, the fluctuating region extended to high above the transition 
temperature\cite{girault}, and 
generated pseudogap behaviour. Its manifestation includes the smooth increase of the spin susceptibility before reaching 
the Pauli limit\cite{johnston1,johnston2}. 

\begin{figure}[h!]
\psfrag{x}[t][b][1.2][0]{$\left(\chi(T)/\chi_0\right)^2$}
\psfrag{y}[b][t][1.2][0]{$R(T)/R_0$}
{\includegraphics[width=7cm,height=7cm]{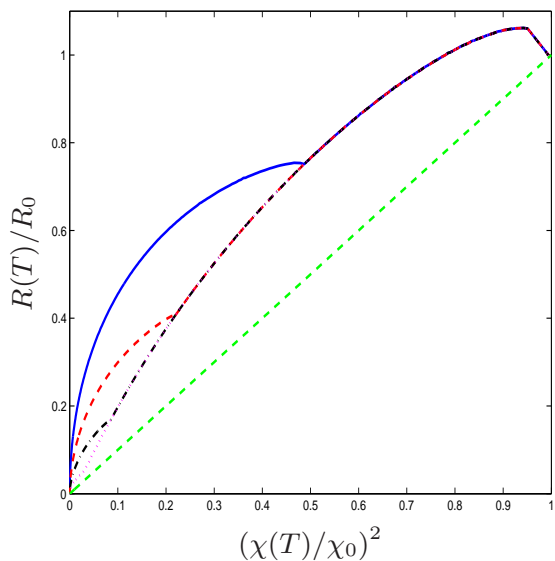}}
\caption{(Color online) The Korringa plot of the relaxation rate for $P_1\rho_0=1$ and
$P_0\rho_0=0.49$ (blue solid line), $0.48$ (red dashed line), $0.47$ (black dashed-dotted line) and $0.46$ (magenta
dotted line). The straight green dashed line is $R(T)/R_0=\left(\chi(T)/\chi_0\right)^2$.
\label{korringa}}
\end{figure}

As opposed to this, (TaSe$_4$)$_2$I exhibits special phenomena. When raising the temperature above the 
charge-density wave transition temperature at 
$T=265$~K, thermal fluctuations disappear in X-ray measurements\cite{requardt}, but the characteristic pseudogap 
behaviour of 
increasing spin susceptibility\cite{johnston2} and spin-lattice relaxation rate\cite{Kriza} remains intact. 
Whence these cannot be related to 
fluctuations. In an effort to resolve this contradiction, we propose an alternative scenario: the material does not 
enter into its 
normal metallic state above $265$~K, but remains in UCDW phase. This is characterized by a gap, which vanishes on 
certain subsets of the Fermi surface. At other points, quasiparticles experience the full energy gap. Fluctuations are 
still present, but their effect is reduced by the available small portion of non-condensed particles ( $50-95$\% of 
the original electrons may belong to the UCDW condensate, as seen in Fig. \ref{spins} through the relation: condensate 
density$=1-\chi(T)/\chi_0$). 
In this respect, our model also accounts for the reduced electronic density of states at the
Fermi energy\cite{schwartz}.
Pseudogap 
behaviour is produced by excitations living around the gap nodes. In the preceding sections, we have successfully 
demonstrated that CDW+UCDW theory reproduces the basic features seen in magnetic probes. The slope of both the spin 
susceptibility and the spin lattice relaxation rate changes when passing through the CDW transition, and both keep on 
increasing with increasing temperature. 
Moreover, a critical divergence of the spin lattice relaxation rate\cite{maniv} at the CDW transition temperature occurs 
usually. As opposed to this,  no sharp feature in $R(T)$ has been detected 
experimentally\cite{Kriza} close to $T_{c0}$, in accordance with our findings, as seen in Fig. \ref{spins}.  
In Fig. \ref{korringa}, we plot the spin-lattice relaxation rate as a function of $\chi^2(T)$. The deviation from the 
Korringa relation, $R(T)/\chi^2(T)=const$ is similar to that found in (TaSe$_4$)$_2$I\cite{Kriza}.

We speculate, that the transition from UCDW to normal metallic state takes place 
above $430$~K, when (TaSe$_4$)$_2$I starts to decompose. As to the conventional CDW transition, it is surely 
overwhelmed by the large phononic contribution in specific heat, although some anomalies are reported\cite{stare}.

As seen in Fig. \ref{phase}, the coexistence is localized into a narrow
region between pure CDW and UCDW. When the system is
forced to leave this regime, it should enter into either a pure CDW or
UCDW. By the application of chemical or mechanical pressure (which, by
increasing the interchain hopping, destabilizes DW phases) or by the
onset of external magnetic field, such a transition can be induced.
The UCDW transition can in this way be brought to the temperature range of feasible
experiments.

Within our weak-coupling mean-field analysis, agreement between theory and experiment is achieved only on qualitative 
levels. The experimental value of the single particle CDW gap is\cite{johnston2} around $1200-1500$~K, which 
implies the gap over transition temperature ratio to be $5-6$, much larger than the weak-coupling BCS value $1.76$. 
We note, that taking imperfect nesting into account\cite{huang1,huang2} would
significantly enhance this ratio\cite{imperfect}, and quantitative agreement could be reached. This is 
 beyond 
the scope of the present paper, but further research in this direction is desirable.

\section{Concluding remarks}

We have studied the coexistence of CDW and UCDW within the mean-field
approximation. Stable coexistence is found only for complex order
parameter, when the relative phase of the CDW and UCDW gaps is $\pm \pi/2$.
By lowering the temperature from the normal state, first UCDW opens at $T_{c1}$.
Than CDW shows up at $T_{c0}$ upon the existing UCDW phase. The CDW (UCDW)
gap increases (decreases) monotonically with decreasing temperature until
$T=0$. Periodic oscillation of the charge density is expected only in
the CDW+UCDW phase.
The opposite scenario, the opening of UCDW gap in a CDW phase has
 not been found in our model.
The density of states possesses a clean gap as long as the CDW phase is
present, and
the square root singularity characteristic to CDW is replaced by the weaker
logarithmic one of UCDW.
The specific heat exhibits two jumps at the corresponding transition
points. The spin susceptibility and spin-lattice relaxation rate increases
steadily without any signs of saturation or coherence peak through
$T_{c0}$ because of the logarithmic nature of the quasiparticle peak in the density of states.

This study is, however, not just of academic interest. Salient features
measured in the pseudogap phase of (TaSe$_4$)$_2$I can be explained with our CDW+UCDW
theory. These are: the increase of both the spin susceptibility and
spin-lattice relaxation rate with temperature above $T_{c0}$, the
featureless behaviour of the latter right below $T_{c0}$, and the rapidly
decaying thermal fluctuations around the CDW transition. 
In this manuscript, we have demonstrated that these phenomena are readily described
by our coexisting CDW and UCDW model. Further studied in this direction are
highly requested in order to reach decisive conclusions.

\begin{acknowledgments}

It is a pleasure to thank L. N\'emeth, Gy. Kriza, K. Maki and P. Matus for instructive discussions.
B. D\'ora was supported by the Magyary Zolt\'an postdoctoral
program of Magyary Zolt\'an Foundation for Higher Education (MZFK).
This work was supported by the Hungarian
Scientific Research Fund under grant numbers OTKA T037976, TS049881, T046269 and NDF45172.
\end{acknowledgments}

\bibliographystyle{apsrev}
\bibliography{eth}
\end{document}